\documentclass[reprint,prl]{revtex4-1}

\usepackage{natbib}
\usepackage{multirow}
\usepackage{amsmath}
\usepackage{graphicx}

\newcommand{\eref}[1]{(\ref{#1})}
\newcommand{\cB}{\mathcal{B}}
\newcommand{\cD}{\mathcal{D}}

\newcommand{\cP}{\mathcal{P}}
\renewcommand{\d}{\mathrm{d}}
\newcommand{\dd}{\mathrm{d}}

\newcommand{\av}[1]{\langle#1\rangle}

\newcommand{\set}[1]{\{#1\}}

\newcommand{\ex}[1]{\e^{#1}}
\newcommand{\e}{\mathrm{e}}
\newcommand{\from}{\leftarrow}

\newcommand{\pij}{p_{ij}}
\newcommand{\pji}{p_{ji}}
\newcommand{\pki}{p_{ki}}
\newcommand{\pkj}{p_{kj}}

\newcommand{\pp}{p_{ij}}
\newcommand{\dist}[1]{\cP(#1)}
\newcommand{\Var}{\Delta^{2}}
\newcommand{\sqVar}{\Delta}

\newcommand{\W}{\widetilde{W}}

\newcommand{\sij}{\sigma_{ij}}
\newcommand{\sji}{\sigma_{ji}}

\newcommand{\ski}{\sigma_{ki}}
\newcommand{\phii}{\phi_{i}}
\newcommand{\phij}{\phi_{j}}
\newcommand{\phik}{\phi_{k}}
\newcommand{\wi}{w_i}
\newcommand{\wj}{w_j}
\newcommand{\wk}{w_k}

\newlength{\figurewidth}
\ifdim\columnwidth<20cm
\else
\fi
\begin{document}
\title{Bayesian Inference of Natural Rankings in Incomplete Competition Networks}
\author{Juyong Park}
\email{Correspondences to \textbf{juyongp@kaist.ac.kr}}
\affiliation{Social Computing Laboratory at the Graduate School of Culture Technology, Korea Advanced Institute of Science and Technology, Daejeon, Republic of Korea 305-701}
\affiliation{Physics Department, Kyung Hee University, Seoul, Republic of Korea 130-701}
\author{Soon-Hyung Yook}
\affiliation{Physics Department, Kyung Hee University, Seoul, Republic of Korea 130-701}

\begin{abstract}
Competition between a complex system's constituents and a corresponding reward mechanism based on it have profound influence on the functioning, stability, and evolution of the system. But determining the dominance hierarchy or ranking among the constituent parts from the strongest to the weakest -- essential in determining reward or penalty -- is almost always an ambiguous task due to the incomplete nature of competition networks. Here we introduce ``Natural Ranking," a desirably unambiguous ranking method applicable to a complete (full) competition network, and formulate an analytical model based on the Bayesian formula  inferring the expected mean and error of the natural ranking of nodes from an incomplete network. We investigate its potential and uses in solving issues in ranking by applying to a real-world competition network of economic and social importance.
\end{abstract}
\maketitle


Understanding of the structure and dynamics of complex networks found in nature, society, and elsewhere have been greatly facilitated by recent advances in the physics of networks~\cite{Newman:2003ia,Barabasi:2005}. Fundamental network problems that have garnered interest include the highly skewed (often power-law) degree (connectivity) distributions, identification of communities or modules in networks, and various critical phenomena and their implications~\cite{Barabasi:1999ef,Newman:2008,Cho08032013}. ``Centrality'' is another concept that is often studied that represents the influence, relevance, or power of a node~\cite{Freeman:1979kg,Newman:2010fk}. Perhaps the best known example is Google's PageRank of webpages, based on a combination of the topology of the hyperlink network and how well the contents of a webpage matches the user search terms (query)~\cite{Page:1998bf}. 

The idea of ranking the nodes based on their relative strengths or relevance -- which we can view as effectively representing competitions between the nodes --  can be useful in many networks; in fact, in many complex systems -- natural, social, or man-made -- the competition-and-reward mechanism is an essential ingredient for their functions and dynamics. Even our daily lives involve continuous decision making based on competitions or comparisons between alternatives in many contexts, ranging from such mundane tasks as choosing where to dine to those very consequential as making critical political or business decisions.

\begin{figure}
\includegraphics[width=80mm]{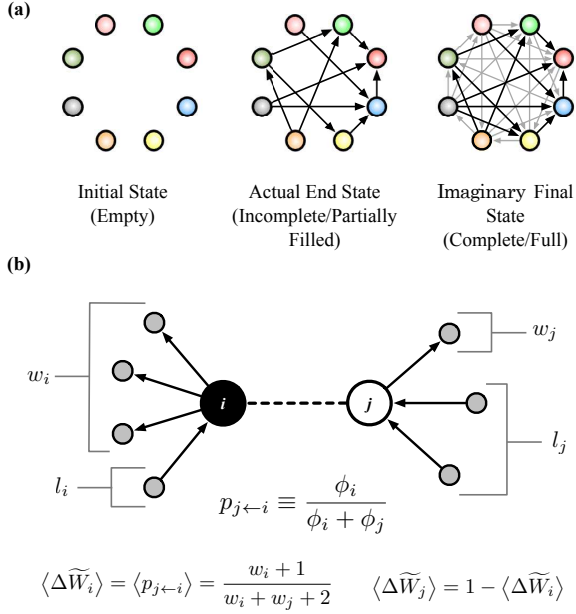}
\caption{(a) An actual incomplete competition network (middle) can be thought of as an intermediary stage of a competition schedule that starts from an empty network (left) and ends as a complete network (right) when all possible competitions have taken place  between nodes. The natural ranking of nodes, applicable in a complete network, is to be estimated (inferred) from the information of wins and losses available at the incomplete stage. (b) The setup of a single-strength parameter model for estimating expected win score of a team from a potential contest. The distribution of strength parameters $\set{\phi}$ can be chosen so that $p_{j\leftarrow i}\equiv\phi_i/(\phi_i+\phi_j)$, the probability that $i$ beats $j$, is fully consistent with Bayes' formula.}
\label{fig01}
\end{figure}

The \textbf{dominance hierarchy} or \textbf{ranking} refers to the linear ordering of things from the strongest to the weakest based on the results of competitions or comparisons. In the case where the thins undergo pairwise (one-to-one) competitions, the entire set of competitions can be represented as a directed network where an arrow points from the winner to the loser of the competition (Fig.~\ref{fig01}~(a)). Food webs in ecological systems (with an arrow pointing from a predator to its prey), sport schedules (with an arrow pointing from the winner to the loser of a game), and merchandise preference testing (with an arrow pointing from the preferred merchandise to those not preferred) are common examples of pairwise competition networks. The dominance hierarchy may take different names -- called the ``trophic level'' in ecology, and ``ranking'' or ``standings'' in sports, for example -- although they are identical. In the remainder of this letter, for convenience we use familiar sports terminology, e.g. ranking, contestant (or player or team), win, loss, tie, and so forth.

A \textbf{complete competition} is one in which every player competes against every body else, also called a round robin~(Fig.~\ref{fig01}~(a)). It would show as a full (complete) network. In such a competition determining the ranking is the easiest: we can simply rank the players in the decreasing order of their total wins $\set{W}$, i.e. the out-degree. When there exists a tie (multiple players with the same $W$), we can employ the following \emph{tie breaker}: We consider the reduced round robin among those tied, and rank them according to their wins therein. This can be applied iteratively to obtain the final ranking in a very simple manner. We call the ranking of nodes obtained this way the ``Natural Ranking,'' as it results from the complete and thus the fairest competition -- every player competes against every other. Note that this is applicable to multiple round-robins as well, as long as each node pair contest an equal number of times. (Note that no tie may be further broken in some cases, for instance when three teams $i$, $j$, and $k$ have the same total wins ($w_i=w_j=w_k$), and $i$ lost to $j$, $j$ lost to $k$, and $k$ lost to $i$, i.e. $\set{\sigma_{ij},\sigma_{jk},\sigma_{ki}}=\set{1,1,1}$ in adjacency matrix notation).

Despite its simplicity and intuitive nature, natural ranking is often inapplicable as many real-world competition networks are incomplete; expecting a real-world competition to be complete is perhaps excessively stringent and unnatural in reality for several reasons. First, the cost of a complete competition can be very high even for moderately large systems -- in a network of $n$ contestants, the number of competitions required is ${n\choose 2}\sim O(n^2)$ -- and thus in the case of the popular US college football of 120 teams, for instance, there may simply be not enough time in a year if one cares for the athletes' health. Second, there may be insurmountable physical constraints as in an ecological food web where the spatial separation between the habitats of two species may hinder them from interacting directly~\cite{Williams:2000oh}.

We can, nevertheless, try to estimate the final natural ranking by imagining that the actual incomplete network we have on hand is merely an intermediary stage of the ``schedule'' of a complete competition that starts from an empty network and ends with a complete one when all competitions have been made (see Fig.~\ref{fig01}~(a)). Then this becomes the problem of inference of a quantity based on currently available information (data), for which the Bayesian framework is one of the most accepted ones in statistics~\cite{MacKay:2003tp}. It can be presented compactly as follows: Labeling $\cP(x)$ the current estimate (also called the Prior) of the distribution of a parameter $x$, and $P(\cD|x)$ the probability of data $\cD$ given $x$ (called the Likelihood), the Bayes formula tells us that one should update $\cP(x)$ via
\begin{align}
	C\cdot P(\cD|x)\cdot\cP(x) \rightarrow \cP(x)
\label{bayesian}
\end{align}
where the new $\cP(x)$ is called the Posterior, and $C$ is the normalization factor so that $\int\cP(x)\d x=1$.

Here we use the Bayesian formula~Eq.~\eref{bayesian} to estimate $\{\widetilde{W}\}$, the projected total win score based on an incomplete competition network to obtain the projected natural ranking. As a generalization of the out-degree, $\widetilde{W}_i$ of node $i$ is the sum of two quantities: The number of actual wins thus far (which we call $w_i$) and the expected number of wins from yet-to-be-played games. Since the latter quantity is equal to the sum of the probabilities of winning the games, our goal becomes estimating $p_{ij}=p_{i\from j}$, our estimation of the probability that $i$ gets defeated by $j$ given the current state of the competition.  To decide $\pij$ consistent with Eq.~\eref{bayesian}, we consider the following.  First, when we have no basis on which to judge the two teams' strengths, e.g. when they have not played any game yet, we are maximally ignorant of $\pij$. This means that it can be any value, i.e. $\cP(\pij)=1$ for $\pij\in[0,1]$~\cite{Jaynes:2003ba,MacKay:2003tp}. Now assume that we observe that $i$ loses to $j$, i.e. we have a datum $\cD=\set{\sij=1}$. Using Eq.~\eref{bayesian} we have the updated $\cP(\pij) = C \pij \cP(\pij)=\pij/2$.  This step offers the foundation for estimating $\cP(\pij)$ between a yet-to-play node pair at any point in the schedule that reflects the strengths of each contestant implied from their performance record.  To achieve this we introduce a strength parameter $\phi_i\in[0,\infty)$ for each contestant such that $\pij$ between two contestants is
\begin{align}
\pij \equiv \frac{\phi_j}{\phi_i+\phi_j}.
\label{defpij}
\end{align}
Using this and the distribution of the strength that we write as $\Phi(\phi)$ we now have
\begin{align}
\dist{\pp}
	&\equiv \int_0^{\infty}\int_0^{\infty}\delta\biggl(\pp-\frac{\phi_j}{\phi_i+\phi_j}\biggr)\Phi(\phi_{i})\Phi(\phi_{j})\d\phi_i\d\phi_j.
\label{fp}
\end{align}
The $\Phi(\phi)$ consistent with Eq.~\eref{bayesian} and Bayes formula can be found as follows. For the flat $\cP(\pij)=1$, we can check using Eq.~\eref{fp} that $\Phi(\phi)=\ex{-\phi}$ for both $\phi_i$ and $\phi_j$. When we observe $\sij=1$, we have $\cP(\pij)=\pij/2$, which is satisfied by the following changes:
\begin{align}
	\Phi(\phi_i) \to \ex{-\phi_i}~~~\textrm{and}~~~\Phi(\phi_j) \to \phi_j\ex{-\phi_j},
\label{newcP}
\end{align}
agreeing with the intuition that $\phi_j$ is likely larger than $\phi_i$ as $\sij=1$ implies that $j$ is likely stronger than $i$. This procedure can be repeated to find a general pattern. Assume now that $j$ (with the one win against $i$) competes against $k$ that has no win, i.e. $\Phi(\phi_k)=\ex{-\phi_k}$. We use $\Phi(\phi_j)$, $\Phi(\phi_k)$, Eqs.~\eref{defpij}~and~\eref{fp} to find the prior $\cP(\pkj)=\pkj/2$ between $j$ and $k$. Then using Eq.~\eref{bayesian} again we have the following two possible updates:
\begin{align}
	\cP(\pkj) \leftarrow \bigl\{\begin{array}{ll} 
		C\cdot\pkj\cdot\pkj = 3\pkj^{2},~\mbox{if $\sigma_{kj}=1$} \\
		C\cdot(1-\pkj)\cdot\pkj = 6(1-\pkj)\pkj,~\mbox{if $\sigma_{jk}=1.$} \nonumber \\
		\end{array}
\label{cPupdatefar}
\end{align}
In a fashion similar to Eq.~\eref{newcP}, the following update rules for$\Phi$s are consistent with Eq.~\eref{cPupdatefar} for \emph{the winner}, while no change is necessary for the loser: 
\begin{align}
	\Phi(\phi_{j}): \phi_{j}\ex{-\phi_{j}} \to \frac{\phi_{j}^{2}\ex{-\phi_{j}}}{2},~\mbox{if $\sigma_{kj}=1$} \nonumber \\
	\Phi(\phi_{k}): \ex{-\phi_{k}} \to \phi_{k}\ex{-\phi_{k}},~\mbox{if $\sigma_{jk}=1.$}
\end{align}
Generally, at a point in the schedule when a player has gathered $w$ wins its $\Phi(\phi)$ is given as
\begin{align}
	\Phi(\phi;w) = \frac{\phi^{w}\ex{-w}}{w!}.
\label{Phiphi}
\end{align}
Using this and Eq.~\eref{fp} the $\cP(\pij)$ between two teams with $w_i$ and $w_j$ actual wins is
\begin{align}
	\cP(\pij)\bigl|_{w_{i},w_{j}} = \frac{\Gamma(w_{i}+w_{j}+2)}{\Gamma(w_{i}+1)\Gamma(w_{j}+1)}(1-\pij)^{w_{i}}\pij^{w_{j}},
\end{align}
from which we have the following simple win score gain for $i$:
\begin{align}
\av{\Delta\widetilde{W}_i}=\av{\pji}=\int_{0}^{1}\pji\cP(\pji)\d\pji=\frac{w_{i}+1}{w_{i}+w_{j}+2}.
\end{align}

Finally, at any given point in the competition, the expected final win score for team $i$ is
\begin{align}
	\widetilde{W}_{i}
		&=\sum_{j\in\Omega_i}\sigma_{ji}+\sum_{j\notin\Omega_i}\av{\pji} \nonumber \\
		&=w_{i}+\sum_{j\notin\Omega_i}\frac{w_{i}+1}{w_{i}+w_{j}+2}
\label{finalscore}
\end{align}
where $\Omega_i$ is the set of nodes that $i$ has competed against. Once a competition becomes complete the second term is zero. In an incomplete competition network, however, the non-zero second term serves as a tiebreaker for teams with the same $w$; an inspection of its functional form tells us that having beaten a stronger opponent counts more than a weaker opponent, which is very intuitive -- -- in sports this is often called the ``strength of schedule''.

Using the exact form for $\Phi(\phi)$, Eq.~\eref{Phiphi}, we can calculate the variance of $\widetilde{W}_{i}$ from
\begin{align}
	\Var\widetilde{W}_{i}
		&= \bigl\langle (W_{i}-\av{W_{i}})^{2}\bigr\rangle \nonumber \\
		&= \sum_{j\notin\Omega_i}(\pji-\pji^{2})+2\sum_{(j<k)\notin\Omega_i}\bigl[\av{\sji\ski}-\pji\pki\bigr],
\end{align}
which needs to be marginalized over $\phi$ in the fashion of Eq.~\eref{fp}. The first part is simple enough:
\begin{align}
	\sum_{j\notin\Omega_i}(\pji-\pji^{2})
		&\to \sum_{j\notin\Omega_i}\frac{1+w_{i}}{2+w_{j}+w_{i}}\biggl(1-\frac{1+w_{i}}{2+w_{j}+w_{i}}\biggr) \nonumber \\
		&= \sum_{j\notin\Omega_i}	\frac{(1+w_{j})(1+w_{i})}{(2+w_{i}+w_{j})^{2}}.
\end{align}
To evaluate the second part, we note that $\av{\sji\ski}\ne\pji\pki=\av{\sji}\av{\ski}$; we say that $\sji$ and $\ski$ are \emph{connected} via $i$, analogous to evaluating Feynman diagrams in field theory. For applications in network theory, see~\cite{Park:2004sm,Park:2010fk}. To evaluate $\av{\sji\ski}$ correctly we need the \emph{joint} probability distribution of $\pji$ and $\pki$ given as
\begin{align}
\cP&(\pji,\pki) \nonumber \\
	&= \int_{\set{\phi}}\delta\biggl(\pji-\frac{\phii}{\phij+\phii}\biggr)\delta\biggl(\pki-\frac{\phii}{\phik+\phii}\biggr) \nonumber \\
	&~~~~~~~~~~\times\Phi(\phii)\Phi(\phij)\Phi(\phik)\dd\phii\dd\phij\dd\phi \nonumber \\
	&= \frac{(\frac{1}{\pji}-1)^{\wj}(\frac{1}{\pki}-1)^{\wk}(\wj+\wk+\wi+2)!}{(\frac{1}{\pji}+\frac{1}{\pki}-1)^{\wj+\wk+\wi+3}\pji^{2}\pki^{2}~\wi!\wj!\wk!},
\label{fjiki}
\end{align}
from which we have
\begin{align}
	\av{\sji\ski}
		&=\int_{0}^{1}\int_{0}^{1}\pji\pki\cP(\pji,\pki)\d\pji\d\pki \nonumber \\
		&\equiv \cB(w_{i},w_{j},w_{k}),
\end{align}
for which we have no closed solution at the time of this writing, although a numerical evaluation is straightforward using symbolic computation packages such as Mathematica. Finally the variance is
\begin{align}
	\Var\widetilde{W}_{i}
		= \sum_{j\notin\Omega_i}&\frac{(1+w_{i})(1+w_{j})}{(2+w_{i}+w_{j})^{2}}+\sum_{(j,k)\notin\Omega_i}\biggl[\cB(\wi,\wj,\wk) \nonumber \\
		&-\frac{(\wi+1)^{2}}{(\wi+\wj+2)(\wi+\wk+2)}\biggr].
\label{variance}
\end{align}

\begin{figure}
\includegraphics[width=80mm]{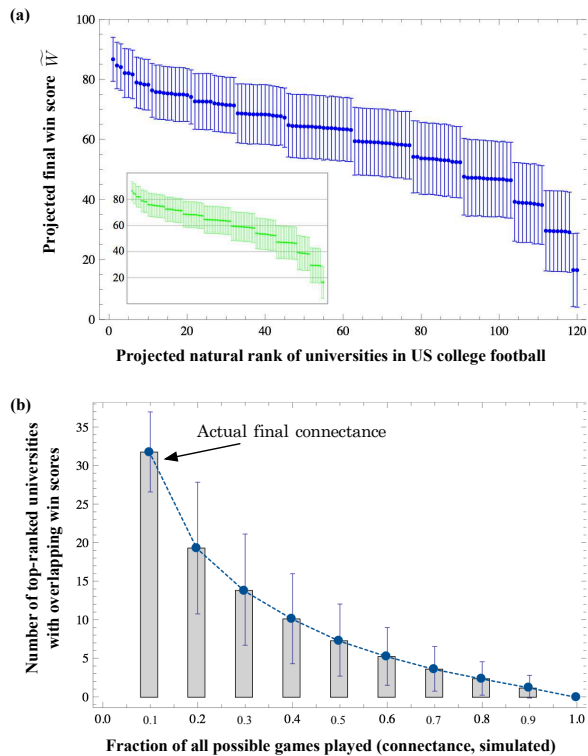}
\caption{(a) The calculated projected final win scores $\set{\widetilde{W}}$ and their mean squared variances $\set{\sqVar\widetilde{W}}$ for the universities that participated the 2010 US football schedule network. The expected scores of universities with identical actual wins are separated by the strength of schedule incorporated in our method. (Inset: Monte Carlo simulation results) (b) The method can be used to estimate the appropriate size of playoff tournaments by investigating the number of highest-ranking teams with overlapping final score ranges in a simulated schedule. A Monte Carlo simulation shows that the network connectance (density) needs to be $\sim 70\%$ for the proposed current four-team playoff system in US college football to be reasonable.}
\label{fig02}
\end{figure}

We now apply our method to US college football to showcase its features and potential. The governing body of the sport called the BCS (short for Bowl Championship Series) which, as in other sports, employs an ``official'' ranking system for the purpose of setting schedules or seeding tournaments~\cite{Stefani:1997bb,Callagan:2004mh,Dunnavant:2004ns}. Given the  popularity of the sport and the substantial benefits -- financial and otherwise -- to successful contestants, the importance of a robust ranking method is essential. Yet the official BCS ranking system, a mixture of human polls and select computer algorithms, is annually an object of outcry from dissatisfied fans. The fundamental origin of the problem is, as mentioned above, the incompleteness of the competition (only $\sim10\%$ of the games are played). Our method applied to this network, presented in Fig.~\ref{fig02}, shows quantitatively the severity of the problem and suggests possible solutions. Fig.~\ref{fig02}~(a) shows the projected win scores $\widetilde{W}$ with the error bars indicating the squared-root--variance $\set{\sqVar W=\Var W^{1/2}}$ as the measure of the uncertainty in $\set{\widetilde{W}}$. First, we note the separation in $\widetilde{W}$ between teams with the same $w$ originating from the strength of schedule in Eq.~\eref{finalscore}, as expected. Also useful for our purposes are  $\set{\sqVar W}$. They clearly show the fundamental limits of the current BCS system that picks the top two teams for the lucrative national championship match: the expected range of $\W$ for the first-ranked team (University of Texas-Austin) overlaps with that of the 30th ranked team; it indicates that the uncertainty is indeed too significant to justify the current BCS method. In 2014 the BCS is poised to adopt a four-team (two-round) playoff tournament to ameliorate the problem, but our results suggest that that too is still insufficient -- in fact, a larger playoff tournament of 32 teams would be more reasonable. Using the fact that the uncertainty decreases as more games are played, we investigated numerically (by creating random schedules beyond what was actually played in the year) the reasonable size of playoff tournaments as the function of the fraction of the games played (i.e. the connectance of the network), shown in Fig.~\ref{fig02}~(b): We see that about 30\% of the possible games need to be played for a sixteen-team playoffs, 50\% for an eight-team playoffs, and 70\%, nearly seven times what is the reality, for the four-team playoffs to be implemented very soon to be reasonable.

In this letter we saw that as a ranking method the natural ranking is attractive as it is intuitive and straightforward, but in many real-life networks it unfortunately cannot be used, as it is applicable only to the rare complete round-robin competitions. In this letter we proposed an analytical model and method that allows us to use the concept of natural ranking in an incomplete network by framing it as an Bayesian inference problem. Starting from the fundamental Bayesian formula Eq.~\eref{bayesian}, we were able to establish a one-parameter model that has incorporates a clear update rule as new information (wins and losses) are uncovered as the competition progresses.  Bayesian inference is fundamentally distribution-based, meaning that it produces not one specific value of a variable but a range of values. This allowed us to estimate not only the mean expectation of the final win scores of teams but their uncertainties (variance), enabling us to answer important questions of practical value, such as the sufficiency of a given playoff system in a major sport, for instance. We hope to see our general method applied to studying various issues in rankings in many complex systems.

We would like to thank Thilo Gross for useful discussions. This work was supported by the National Research Foundation of Korea (Grant NRF-20100004910), Korea Advanced Institute of Science and Technology, and Kyung Hee University (Grant KHU-201020100116).


\bibliographystyle{apsrev}
\bibliography{references}

\end{document}